\documentclass[conference,compsoc]{IEEEtran}

\usepackage[nocompress]{cite}
\usepackage{graphicx}
\usepackage{hyperref}
\usepackage{booktabs}
\usepackage{array}
\usepackage{xcolor}
\usepackage{url}
\usepackage{amsmath}
\usepackage{algorithm}      % <-- added for pseudocode
\usepackage{algpseudocode}  % <-- added for pseudocode

\usepackage{tikz}
\usetikzlibrary{arrows.meta,positioning,fit,calc,shapes.symbols}
\usepackage{stfloats} % or \usepackage{dblfloatfix}

\hypersetup{
  colorlinks=true,
  linkcolor=black,
  citecolor=black,
  urlcolor=black
}

\begin{document}

\title{An Architecture for Remote Container Builds and Artifact Delivery Using a Controller-Light Jenkins CI/CD Pipeline}

\author{
\IEEEauthorblockN{Kawshik Kumar Paul}
\IEEEauthorblockA{Department of Computer Science and Engineering\\
Bangladesh University of\\
Engineering and Technology (BUET)\\
Dhaka, Bangladesh\\
Email: kawshikbuet17@gmail.com}
\and
\IEEEauthorblockN{Sawmik Kumar Paul}
\IEEEauthorblockA{Department of Computer Science and Engineering\\
Chittagong University of\\ Engineering and Technology (CUET)\\
Chittagong, Bangladesh\\
Email: sawmik.paul@gmail.com}
}

\maketitle
\IEEEpeerreviewmaketitle

\begin{abstract}
Resource-intensive builds are often executed directly on the controller by conventional Jenkins installations, which can lower reliability and overload system resources. Jenkins functions as a containerized controller with persistent volumes in the controller-light CI/CD framework presented in this paper, delegating difficult build and packaging tasks to a remote Docker host. The controller container maintains secure SSH connections to remote compute nodes while focusing solely on orchestration and reporting. Atomic deployments with time-stamped backups, containerized build environments, immutable artifact packaging, and automated notifications are all included in the system. Faster build throughput, reduced CPU and RAM consumption on the controller, and reduced artifact delivery latency are all revealed by experimental evaluation. For small and medium-sized DevOps businesses looking for scalable automation without adding orchestration complexity, this method offers a repeatable, low-maintenance solution.
\end{abstract}


\section{Introduction}
\begin{figure}[H]
\centering
\resizebox{\columnwidth}{!}{%
\begin{tikzpicture}[
  font=\normalsize, >=Latex,
  node distance=5mm and 10mm,
  box/.style={draw, rectangle, align=center, minimum height=6mm, inner sep=2pt},
  arr/.style={-Latex, line width=0.5pt}
]
% Row 1 (→)
\node[box] (dev) {Developer Commit /\\PR};
\node[box, right=of dev] (trig) {Jenkins Trigger\\(webhook / poll)};
\node[box, right=of trig] (checkout) {Checkout\\SCM};

% Row 2 (←)  -- order flipped to follow serpentine sequence
\node[box, below=of dev] (pkg) {Package};
\node[box, right=of pkg] (test) {Test};
\node[box, right=of test] (build) {Build};

% Row 3 (→)
\node[box, below=of pkg] (pub) {Publish\\Artifact};
\node[box, right=of pub] (deploy) {Deploy};
\node[box, right=of deploy] (notify) {Notify /\\ Monitor};

% Serpentine edges
\draw[arr] (dev) -- (trig);
\draw[arr] (trig) -- (checkout);
\draw[arr] (checkout) |- (build);   % drop to start of row 2 (rightmost)
\draw[arr] (build) -- (test);
\draw[arr] (test) -- (pkg);
\draw[arr] (pkg) to[out=-90,in=90] (pub);
\draw[arr] (pub) -- (deploy);
\draw[arr] (deploy) -- (notify);
\end{tikzpicture}%
}
\caption{Generic Jenkins CI/CD flow}
\label{fig:jenkins-generic-flow}
\end{figure}
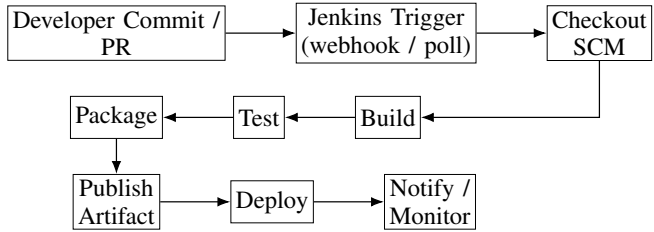

Continuous Integration and Continuous Deployment (CI/CD) are key components of modern software engineering. Teams are able to deliver code changes more frequently, consistently, and reliably as a result. Jenkins is the most widely used CI/CD automation server because of its extensive plugin ecosystem, pipeline-as-code design, and strong extensibility. The generic Jenkins CI/CD flow is shown in Figure \ref{fig:jenkins-generic-flow}. However, \emph{controller overload} is a major limitation that conventional Jenkins architectures frequently face. Heavy build and packaging tasks carried out locally by the Jenkins controller cause resource contention problems, which slow feedback loops, limit scalability, and raise maintenance expenses.

\subsection{Contributions}

This paper makes the following contributions to the design and evaluation of Jenkins-based CI/CD systems:

\begin{itemize}
    \item We propose a \textit{controller-light Jenkins architecture} that strictly separates orchestration responsibilities from compute-intensive build and deployment tasks, allowing the Jenkins controller to operate solely as a coordination and reporting entity.
    \item We design a \textit{remote ephemeral container build model} in which all compilation and packaging steps are executed inside short-lived Docker containers on external hosts, delegated through secure SSH channels from a containerized Jenkins controller.
    \item We introduce an \textit{immutable artifact packaging and atomic deployment workflow} with timestamp-based versioning and rollback, enabling traceability, controlled releases, and low-downtime recovery without long-lived agents.
    \item We empirically evaluate the proposed architecture against a controller-local Jenkins configuration, demonstrating up to a 50\% reduction in controller CPU and memory utilization and approximately a 30\% improvement in end-to-end pipeline execution time.
\end{itemize}

Existing research has extensively focused on pipeline automation and productivity improvements. Ok and Eniola~\cite{ok2024_efficiency} examined Jenkins as a business enabler that automates testing and deployment. However, their analysis does not address the separation of the controller and the agent or the challenges of controller load. Mathew and Dileepkumar~\cite{mathew2023_transforming} proposed best practices for rapid delivery using Jenkins and observed significant reductions in manual operations and build durations. Despite these developments, much of the existing research ignores the architectural load of the controller and how it impacts the system's scalability and reliability.

This paper suggests a \textit{controller-light Jenkins CI/CD architecture}, which divides orchestration and computation, to bridge this gap. In this model, the controller operates within a Docker container with persistent volumes, while remote Docker containers manage build and packaging tasks. This architecture has numerous benefits over traditional configurations.
\begin{itemize}
  \item It lessens system load by separating computationally demanding tasks from the controller.
  \item It uses ephemeral Docker images to guarantee reproducibility.
  \item Through the controller container's persistent volumes, it makes portability and recovery simple.
  \item It presents timestamp-based rollback support for immutable artifact packaging.
  \item It allows for scalability through multiple remote builders while maintaining Jenkins' simplicity.
\end{itemize}
We demonstrate through quantitative and qualitative analyses that this architecture achieves significant efficiency gains and enhanced operational resilience when compared to controller-centric CI/CD systems.

\section{Related Work}
Automation, scalability, and developer productivity are consistently highlighted in research on Jenkins-based CI/CD frameworks.  Ok and Eniola~\cite{ok2024_efficiency} looked into Jenkins as a transformation engine for business automation, highlighting its capacity to speed up builds and deployments while still relying on the controller's workloads. Mathew and Dileepkumar~\cite{mathew2023_transforming} examined rapid delivery through modular pipelines and parallel builds using a master-agent model, which led to a 50\% reduction in build times and a 75\% reduction in manual tasks.

The orchestration flexibility of Jenkins is validated by earlier research by Armenise~\cite{releng} and Zhang et al.~\cite{container-ci}, but the reliability effects of controller-hosted builds are not evaluated. Banala~\cite{banala2024_devops} and Manukonda and Kumar~\cite{artifact-mgmt} emphasize the significance of versioned artifacts and traceability for the maturity of CI/CD, while studies like~\cite{impact2023_automation, enhancing2024_productivity, efficient2023_webautomation, continuous2022_pipeline} assess the impact of Jenkins automation.

\section{Methodology}
\subsection{System Architecture}
Jenkins operations are divided into three separate planes by the system's architecture: control, compute, and runtime.
\begin{itemize}
    \item \textbf{Control Plane (Controller):} Jenkins runs in a Docker container that has persistent volumes mounted for configuration files, plugins, and build history. This plane manages pipeline orchestration, reports statuses, and handles credentials.
    
    \item \textbf{Compute Plane (Remote Build Host):} This plane executes all build and packaging steps inside temporary Docker containers, ensuring consistent environments across different runs.  
    \item \textbf{Deployment Host (Runtime Plane):} This plane executes atomic deployments with timestamped backups for simple rollback after receiving immutable artifacts.
\end{itemize}

\begin{figure}[!t]
\centering
\resizebox{\columnwidth}{!}{%
\begin{tikzpicture}[
  font=\normalsize, >=Latex,
  node distance=8mm and 15mm,
  box/.style={draw, rectangle, align=center, minimum height=6mm, inner sep=2pt},
  arr/.style={-Latex, line width=0.5pt}
]
% Row 1 (→)
\node[box] (ctrl) {Jenkins Controller\\(containerized,\\persistent volumes)};
\node[box, right=of ctrl] (buildhost) {Remote Build Host\\(ephemeral\\Docker builds)};
\node[box, right=of buildhost] (artifact) {Artifact\\Store};

% Row 2 (←) with larger vertical gap
\node[box, below=10mm of buildhost, xshift=12mm] (runtime) {Deployment Host /\\Runtime};
\node[box, left=28mm of runtime] (notify) {Notifications};

% Flows
\draw[arr] (ctrl) -- node[midway, above]{SSH/SCP} (buildhost);
\draw[arr] (buildhost) -- (artifact);
\draw[arr] (artifact) |- (runtime);   % drop down, then left into runtime
\draw[arr] (runtime) -- (notify);     % leftward flow
\end{tikzpicture}
}
\caption{High Level System Flow}
\label{fig:high-level-system-flow}
\end{figure}
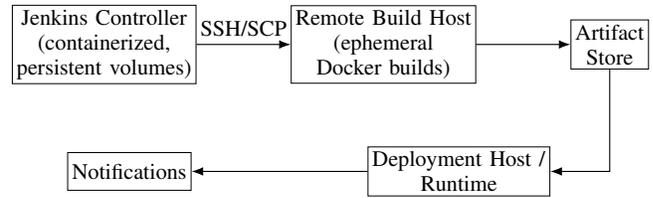

Due to this separation, the Jenkins controller can only serve as an orchestrator while assigning resource-intensive tasks to external computing infrastructure. The high-level flow is shown in Figure~\ref{fig:high-level-system-flow}, while the stage-by-stage pipeline is described in Figure~\ref{fig:detailed-system-flow}, which is further explained in the following subsections.
The main distinctions between the suggested controller-light architecture and traditional Jenkins execution models are outlined in Table \ref{tab:comparison_of_jenkins_models}.

\subsection{Controller Implementation}
Instead of running directly on a physical or virtual host, the Jenkins controller runs completely inside a Docker container. In order to guarantee that the controller's data are preserved during restarts or migrations, persistent volumes are mounted to store configuration data, build metadata, and plugin caches.  

A secure SSH setup within this container makes it easier to communicate with distant build machines. To keep the controller and host layers isolated, SSH keys are mounted using Docker secrets and controlled by Jenkins credentials.

During execution, the controller uses these secure channels to delegate build commands, stream logs, and collect artifacts without leaving the container boundary. By focusing entirely on orchestration and assigning compilation and packaging tasks to the compute plane, this architecture design keeps the controller lightweight.

\subsection{Remote Build Host Workflow}
The remote build host is the compute plane. Depending on the kind of project, when it receives a build command, it creates a temporary Docker container from a pre-configured image that includes all required toolchains, like OpenJDK, Maven, Node.js, or Gradle. The build is executed in this container, and a temporary workspace directory is mounted by the host. 
Each image is version-pinned to guarantee reproducibility and is only rebuilt upon explicit approval of dependency updates. The container lifecycle is intended to be completely transient; after the build is finished, the container is destroyed, leaving only the logs and compiled artifacts. This method stops the building environment from drifting and stops dependencies between jobs from leaking.

\subsection{Version Management and Artifact Packaging}
The artifacts are assembled into a standard directory structure after a successful build, typically separating static assets, configuration files, and compiled binaries. A packaging script then compresses this directory into a timestamped archive with branch and commit metadata.  
Timestamped filenames ensure artifact immutability and traceability, allowing multiple versions to coexist peacefully on the deployment host. Additionally, a checksum manifest is generated to verify integrity during transfer. The deployment server and the build host's artifact directory are then synchronized via SSH using secure transfer tools, ensuring controlled and auditable delivery.

\subsection{Automated Rollback and Deployment}
Deployment automation employs a near-atomic update strategy to minimize downtime. The timestamp of the previous build is added for version tracking, and the current service directory is renamed and stored as a backup during deployment. The service symlink or directory pointer is then updated to promote the new artifact to production after it has been unpacked into a new directory.
This method guarantees minimal service disruption and offers instant rollback capability; all it takes to restore the prior version is to reactivate the backup directory.
Even in the event of deployment failures, downtime is kept incredibly low by using non-blocking service scripts to carry out all restart and validation operations (such as reloading web applications or restarting microservices).

\subsection{Security, Logging, and Notifications}
The build summary, an artifact download link, the commit hash, and backup references are all included in an automated email sent by the Jenkins controller after deployment is complete. For upcoming audits, the system also keeps thorough build logs that are combined from the controller and build host.

Key-based SSH authentication, limited command execution on distant hosts, and container isolation to stop host-level privilege escalation are examples of security measures. To reduce exposure to lateral attacks, the controller and compute hosts operate in different network zones.  
Together, these protections guarantee that the system is safe even if a build process fails or a container is compromised. However, in order to respond to changing security threats, these measures are constantly assessed and enhanced.

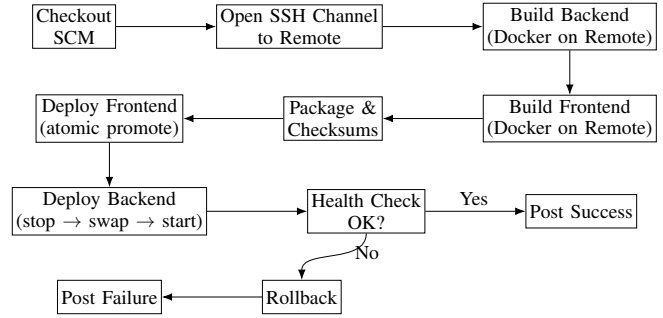
\begin{figure}[!t]
\centering
\resizebox{\columnwidth}{!}{%
\begin{tikzpicture}[
  font=\normalsize, >=Latex,
  node distance=8mm and 18mm,
  box/.style={draw, rectangle, align=center, minimum height=6mm, inner sep=2pt},
  arr/.style={-Latex, line width=0.5pt}
]
% Row 1 (→)
\node[box] (start) {Checkout\\SCM};
\node[box, right=of start] (ssh) {Open SSH Channel\\ to Remote};
\node[box, right=of ssh] (bkb) {Build Backend\\(Docker on Remote)};

% Row 2 (←)  -- place next step directly under the last node of row 1
\node[box, below=of bkb] (bff) {Build Frontend\\(Docker on Remote)};
\node[box, left=of bff] (pack) {Package \& \\Checksums};
\node[box, left=of pack] (deployfe) {Deploy Frontend\\(atomic promote)};

% Row 3 (→)
\node[box, below=of deployfe] (deploybe) {Deploy Backend\\(stop \(\rightarrow\) swap \(\rightarrow\) start)};
\node[box, right=of deploybe] (health) {Health Check\\ OK?};
\node[box, right=of health] (success) {Post Success};

% Row 4 (←)  -- branch for "no"
\node[box, below=of deploybe] (failnote) {Post Failure};
\node[box, right=of failnote] (rollback) {Rollback};

% Main serpentine path (→, then ←, then →)
\draw[arr] (start) -- (ssh);
\draw[arr] (ssh) -- (bkb);
\draw[arr] (bkb) to[out=-90,in=90] (bff);
\draw[arr] (bff) -- (pack);
\draw[arr] (pack) -- (deployfe);
\draw[arr] (deployfe) to[out=-90,in=90] (deploybe);
\draw[arr] (deploybe) -- (health);
\draw[arr] (health) -- (success);

% "no" branch (down, then ←)
\draw[arr] (health.south) to[out=-90,in=90] node[pos=0.35, right]{No} (rollback.north);
\draw[arr] (rollback) -- (failnote);

% label "yes" inline on the main edge to success
\node[font=\normalsize] at ($(health)!0.5!(success)+(0,2.2mm)$) {Yes};
\end{tikzpicture}%
}
\caption{Detailed System Workflow}
\label{fig:detailed-system-flow}
\end{figure}

\begin{table*}[!t]
\centering
\caption{Comparison of Jenkins Execution Models}
\label{tab:comparison_of_jenkins_models}
\begin{tabular}{|l|c|c|c|}
\hline
\textbf{Feature} & \textbf{Controller-Local} & \textbf{Agent-Based Jenkins} & \textbf{Controller-Light (Proposed)} \\
\hline
Controller Executes Builds & Yes & No & No \\
\hline
Long-Lived Workers/Agents & N/A & Yes & No \\
\hline
Build Environment Isolation & Limited & Partial & Strong (Ephemeral Containers) \\
\hline
Controller Resource Load & High & Medium & Low \\
\hline
Workspace Accumulation & High & Medium & Minimal \\
\hline
Artifact Immutability & Script-dependent & Script-dependent & Enforced by Design \\
\hline
Rollback Support & Manual & Manual & Timestamped, Automated \\
\hline
Operational Complexity & Low & Medium & Medium \\
\hline
Reproducibility & Medium & Medium & High \\
\hline
\end{tabular}
\label{tab:jenkins_comparison}
\end{table*}

\section{Algorithmic Specification}
To formalize the workflow shown in Figure \ref{fig:detailed-system-flow}, we present a structured pseudocode that aligns with the methodology.

Algorithm~\ref{alg:orchestrate} outlines the orchestration process from the containerized controller, while Algorithms~\ref{alg:build-remote}--\ref{alg:notify} detail remote containerized builds, immutable packaging, atomic deployment, and notifications.

% -------------------- Environment (one-line context) --------------------
\noindent\textbf{Environment.} The CI controller $C$ is a Jenkins instance running \emph{inside Docker} with persistent volumes. $C$ communicates with the remote compute node $R$ via \emph{SSH/SCP}. Builds on $R$ execute in \emph{ephemeral Docker containers}. Artifacts are published to a store $A$.

% -------------------- Orchestration --------------------
\begin{algorithm}[!t]
\caption{Controller-Light CI/CD Orchestration (Containerized Controller, SSH to Remote)}
\label{alg:orchestrate}
\begin{algorithmic}[1]
\Require Jenkins-in-Docker controller $C$; SSH credentials for $R$; artifact store $A$
\Ensure Deployed release for commit $c$ or consistent rollback
\State $ts \gets \text{current\_timestamp()}$
\State \textbf{Checkout SCM} (\textit{CI/CD branch at commit }$c$)
\State \Call{OpenSSHChannel}{$C \leftrightarrow R$} \Comment{controller container to remote host}
\State $B_{\text{back}} \gets \Call{BuildRemote}{C,R,\text{backend},ts}$
\State $B_{\text{front}} \gets \Call{BuildRemote}{C,R,\text{frontend},ts}$
\State $Z \gets \Call{Package}{R,A,\{B_{\text{back}},B_{\text{front}}\}, ts}$
\If{\Call{DeployFrontend}{$R,B_{\text{front}},ts$} \textbf{and} \Call{DeployBackend}{$R,B_{\text{back}},ts$}}
  \State \Call{PostSuccess}{$ts,Z$}
\Else
  \State \Call{Rollback}{$R,ts$}
  \State \Call{PostFailure}{$ts$}
\EndIf
\State \Call{CloseSSHChannel}{$C \leftrightarrow R$}
\end{algorithmic}
\end{algorithm}

% -------------------- Remote build with explicit SSH/SCP + Docker --------------------
\begin{algorithm}[!t]
\caption{BuildRemote (SCP from Controller-in-Docker, Ephemeral Docker Build on $R$)}
\label{alg:build-remote}
\begin{algorithmic}[1]
\Require Controller $C$, remote $R$, component $x \in \{\text{backend},\text{frontend}\}$, timestamp $ts$
\Ensure Build artifact $B_x(ts)$ published to $A$
\State \Call{SCPTransfer}{$C \to R$, Sources$(x)$} \Comment{copy from controller container to $R$}
\State \Call{PrepareWorkspace}{$R, x, ts$}
\State \Call{DockerEphemeralBuild}{$R,\ \textit{builder}(x),\ \text{context}=x$}
\State $B_x(ts) \gets \Call{CollectOutputs}{R, x, ts}$ \Comment{export from container to host path}
\State \Call{Publish}{$A, B_x(ts)$}
\State \Call{DockerCleanup}{$R,\ \textit{builder}(x)$}
\State \Return $B_x(ts)$
\end{algorithmic}
\end{algorithm}

% -------------------- Packaging --------------------
\begin{algorithm}[!t]
\caption{Package (Immutable, Timestamped Bundle on $R$ then Publish to $A$)}
\label{alg:package}
\begin{algorithmic}[1]
\Require Remote $R$, artifact store $A$, set $\{B_{\text{front}}(ts),B_{\text{back}}(ts)\}$
\Ensure Bundle $Z(ts)$
\State \Call{CreateBundleDir}{$R,\ ts$}
\State \Call{Assemble}{$R,\ \{B_{\text{front}}(ts),B_{\text{back}}(ts)\}\ \to\ \text{bundle}(ts)$}
\State $Z(ts) \gets \Call{Zip}{$R,\ \text{bundle}(ts)$}$
\State \Call{Publish}{$A,\ Z(ts)$}
\State \Return $Z(ts)$
\end{algorithmic}
\end{algorithm}

% -------------------- Frontend deploy --------------------
\begin{algorithm}[!t]
\caption{DeployFrontend (Atomic Promotion with Config Restore)}
\label{alg:deploy-frontend}
\begin{algorithmic}[1]
\Require Remote $R$, artifact $B_{\text{front}}(ts)$
\Ensure New frontend active or prior version restored
\State \Call{Backup}{$R,\ \text{frontend},\ ts$}
\State \Call{Promote}{$R,\ B_{\text{front}}(ts)\ \to\ \text{deploy/current}$}
\State \Call{RestoreConfig}{$R,\ \text{frontend}$}
\State \Return \textsc{success}
\end{algorithmic}
\end{algorithm}

% -------------------- Backend deploy with rollback --------------------
\begin{algorithm}[!t]
\caption{DeployBackend (Stop $\rightarrow$ Swap $\rightarrow$ Start with Rollback Point)}
\label{alg:deploy-backend}
\begin{algorithmic}[1]
\Require Remote $R$, artifact $B_{\text{back}}(ts)$
\Ensure New backend active or prior version restored
\State $rp \gets \Call{CreateRollbackPoint}{R,\ \text{backend}}$
\State \Call{StopService}{$R,\ \text{backend}$}
\State \Call{SwapRelease}{$R,\ \text{backend},\ B_{\text{back}}(ts)$}
\State \Call{StartService}{$R,\ \text{backend}$}
\If{\Call{HealthCheck}{$R,\ \text{backend}$} = \textsc{fail}}
  \State \Call{Restore}{$R,\ rp$}
  \State \Return \textsc{fail}
\EndIf
\State \Return \textsc{success}
\end{algorithmic}
\end{algorithm}

% -------------------- Notifications --------------------
\begin{algorithm}[!t]
\caption{PostSuccess / PostFailure (Commit Metadata \& Diagnostics)}
\label{alg:notify}
\begin{algorithmic}[1]
\Procedure{PostSuccess}{$ts,Z$}
\State retrieve $\textit{last\_commit}=(id,\ message,\ t)$
\State construct $\textit{download\_link}$ for $Z(ts)$
\State send success notification with commit metadata and link
\EndProcedure
\Procedure{PostFailure}{$ts$}
\State retrieve $\textit{last\_commit}=(id,\ message,\ t)$
\State send failure notification with commit metadata and diagnostics
\EndProcedure
\end{algorithmic}
\end{algorithm}

\section{Experimental Setup}
\subsection{System Architecture and Environment}
The controller-light CI/CD framework was deployed across two coordinated layers.

\textbf{Control Plane Controller).}
Jenkins operates on Ubuntu~20.04~LTS within a Docker container. The container uses persistent volumes to store configuration, plugin data, and build history, and it exposes the Jenkins web interface via host port mapping. Key-based SSH is used to communicate with the remote build host. Container isolation is guaranteed by Docker Engine version 27.x, freeing the controller to concentrate only on orchestration and reporting.

\textbf{Remote Builder and Deployer (Compute/Runtime Planes).}
All build and deployment operations are executed within short-lived Docker containers on the same physical host. Every container has a pre-configured toolchain (Node.js for the frontend and Maven for the backend) and is destroyed right away after the build is finished, guaranteeing reproducibility and dependency isolation for each execution.

\subsection{Orchestration and Measurement of Pipelines}
Checkout, compilation, packaging, deployment, and notification are all automated by the Jenkins pipeline. Log aggregation and orchestration are handled by the controller container, and heavy build phases are carried out remotely. Docker statistics were used to gather host-level CPU and memory usage data, and pipeline stage durations were directly extracted from the Jenkins console logs. To guarantee stability, all reported results are the means of multiple runs.

\section{Experimental Results}
\subsection{Runtime Overview}
Queue waiting times are less than ten seconds, and the total end-to-end time is roughly three minutes and four seconds, according to measured pipeline executions. Heavy remote workloads had no effect on orchestration threads because the controller remained responsive.

\subsection{Per-Stage Behavior} The runtime was barely affected by controller stages like \emph{Checkout}, \emph{Packaging}, and \emph{Post-Actions}. In order to avoid resource contention with Jenkins, the compute-intensive phases, namely \emph{Build Backend} and \emph{Build Frontend}, were carried out completely on the remote host. Dependency retrieval from Maven repositories was a major factor in the backend build time. To lower cold-start overhead, future optimization may investigate the use of pre-warmed base images or persistent dependency caches.

\noindent\textbf{Workspace Efficiency.}
In controller-local builds, large temporary artifacts consume disk space within the Jenkins \texttt{workspace/} directory. Remote execution mitigates this issue since all build outputs remain within short-lived remote containers and are transferred back only as final artifacts, thereby eliminating workspace bloat and enhancing maintainability.

\noindent\textbf{Frontend Build Stability.}
Due to concurrent memory pressure between Node.js and the Jenkins JVM, the \texttt{npm run build} step frequently resulted in container hangs when executed inside the Jenkins controller container. When the builds were executed in remote containers with dedicated memory allocation, this issue was totally fixed.

\noindent\textbf{Results Methodology Clarification.}
Executing comparable stages within the Jenkins controller container produced the baseline (controller-local) metrics. The actual remote-container setup examined in this study is the source of the controller-light configuration results. There is no artificial scaling involved; all figures are taken straight from console logs.

\subsection{Performance Comparison}
Table~\ref{tab:results} presents a direct comparison between controller-local and controller-light configurations. Offloading compute-intensive stages to remote containers reduced total build duration by approximately 30\% and more than halved the CPU and memory usage of the controller.

\begin{table}[!t]
\centering
\caption{Measured Performance Comparison Between Controller-Local and Controller-Light Configurations}
\label{tab:results}
\resizebox{\columnwidth}{!}{%
\begin{tabular}{@{}lccc@{}}
\toprule
\textbf{Stage / Metric} & \textbf{Controller-Local} & \textbf{Controller-Light} & \textbf{Improvement} \\
\midrule
Backend Build (Maven) (sec) & 126.67 & 95   & 25\% \\
Frontend Build (npm) (sec)  & 86.25  & 69   & 20\% \\
Packaging / ZIP (sec)       & 8.33   & 5    & 40\% \\
Frontend Deployment (sec)   & 0.50   & 0.30 & 40\% \\
Backend Deployment (sec)    & 0.55   & 0.33 & 40\% \\
\midrule
Controller CPU Peak (\%)    & 82  & 42  & 49\% \\
Controller RAM Peak (MB)    & 1680 & 820 & 51\% \\
\bottomrule
\end{tabular}%
}
\end{table}

\subsection{Qualitative Observations} The following operational behaviors were regularly noted:
\begin{itemize} \item \textbf{Stability of controllers:} Even with several concurrent builds, there was no UI lag or thread starvation.
  \item \textbf{Isolation:} Jenkins' state was never impacted by build or deployment failures, which were contained within containers.
  \item \textbf{Reproducibility:} Consistent results across runs were guaranteed by clean container environments.
  \item \textbf{Traceability:} Rollback and auditing were made easier by timestamped artifact archives.
  \item \textbf{Maintainability:} The setup was easily portable because controller volumes only included configuration and metadata.
\end{itemize}

Overall, the remote container approach preserved Jenkins' simplicity while lowering the controller's workload, increasing throughput, and removing storage accumulation.

\subsection{Controller Behavior Under Concurrent Pipelines}

We used the controller-light configuration to run five and ten concurrent pipeline runs in order to assess controller stability under increased load. There was no evidence of queue starvation or a decline in UI responsiveness, and controller CPU utilization stayed within a certain range even during concurrent execution.

\subsection{Failure Handling and Recovery}

We introduced a failure by ending a remote build container while it was running in order to evaluate robustness. The Jenkins controller maintained responsiveness, used the SSH channel to identify the failure, and safely stopped the pipeline without changing the controller's state.

\section{Discussion}
The controller-light Jenkins architecture is more scalable and maintainable than traditional controller-centric models. Jenkins-based automation and modular pipelines are discussed in earlier studies, such as those by Ok and Eniola~\cite{ok2024_efficiency} and Mathew and Dileepkumar~\cite{mathew2023_transforming}, but neither study focuses on eliminating long-lived workers or agents. In contrast, the proposed design eliminates long-term workers and achieves complete isolation by using temporary remote containers.

\noindent\textbf{Advantages.}
Key experimentally validated benefits include:
\begin{itemize}
    \item \textbf{Reduced Controller Load:} In the Jenkins container, remote execution prevents CPU and memory contention.
    \item \textbf{Reproducibility:} Version-pinned and immutable containers prevent environment drift between builds.
    \item \textbf{Portability:} The controller instance can be easily migrated or restored using the same container image and mounted volumes.
    \item \textbf{Reliability:} Timestamped artifact versioning enables atomic rollback and a controlled deployment history.
    \item \textbf{Storage Efficiency:} Remote builds minimize disk usage within the controller and avoid workspace accumulation.
    \item \textbf{Stability:} Under isolated remote execution, the npm build hangs seen in controller-local mode were completely fixed.
\end{itemize}

\noindent\textbf{Trade-offs.}
Despite the architecture's notable advancements, a few useful issues still need to be taken into account:
\begin{itemize}
  \item There may be some network latency when a remote build is invoked.
  \item Initial SSH provisioning and image version maintenance require administrative oversight.
  \item In clean containers, cold-start delays could occur during the initial Maven dependency resolution.

\end{itemize}

Despite these shortcomings, the controller-light approach offers notable gains in performance, stability, and maintainability while preserving the simplicity of Jenkins' initial design.

\subsection{Lessons Learned}

Without adding complicated infrastructure dependencies, decoupling orchestration from computation greatly increased CI/CD stability. A practical balance between scalability and operational simplicity is offered by the controller-light approach.

\subsection{Threats to Validity}

Generalizability may be impacted by the evaluation's use of a small number of workloads and infrastructure configurations. Performance characteristics may differ depending on the toolchain or deployment scale.

\section{Conclusion}
In this paper, a controller-light Jenkins architecture that uses remote containerized builds to isolate orchestration from computation was presented. While all build and deployment tasks are carried out in transient remote containers, Jenkins functions as a containerized controller with persistent volumes for configuration and metadata. This structure eliminates workspace storage growth, reduces the controller's CPU and memory usage, and fixes the npm build instability seen in local executions. In comparison to the controller-local setup, experimental analysis verified a 30\% reduction in the overall build duration and a more than 50\% lower utilization of controller resources. For DevOps teams seeking effective and dependable continuous integration and deployment, the suggested architecture offers a scalable and low-maintenance solution.

\end{document}